\documentclass[english,aps]{revtex4-2}
\usepackage[T1]{fontenc}
\usepackage[latin9]{inputenc}
\usepackage{amsmath}
\usepackage{graphicx}
\usepackage{babel}

\begin{document}
\title{Anomalous diffusion in nonlinear transformations of the noisy voter
model }
\author{Rytis Kazakevi\v{c}ius}
\email{rytis.kazakevicius@tfai.vu.lt}
\affiliation{Institute of Theoretical Physics and Astronomy, Vilnius University,
Saul\.{e}tekio 3, LT-10257 Vilnius, Lithuania}
\author{Aleksejus Kononovicius}
\email{aleksejus.kononovicius@tfai.vu.lt}
\homepage{http://kononovicius.lt}
\affiliation{Institute of Theoretical Physics and Astronomy, Vilnius University,
Saul\.{e}tekio 3, LT-10257 Vilnius, Lithuania}
\begin{abstract}
Voter models are well known in the interdisciplinary community, yet
they haven't been studied from the perspective of anomalous diffusion.
In this paper we show that the original voter model exhibits ballistic
regime. Nonlinear transformations of the observation variable and
time scale allows us to observe other regimes of anomalous diffusion
as well as normal diffusion. We show that numerical simulation results
coincide with derived analytical approximations describing the temporal
evolution of the raw moments.
\end{abstract}
\maketitle

\section{Introduction}

There is a broad range of systems where the time dependence of the
centered second moment is not linear as in the classical Brownian
motion. Such a family of processes is called anomalous diffusion. In
one dimension the anomalous diffusion is defined by the power law
time dependence of the mean square displacement (MSD)
$\left\langle \left(\Delta x\right)^{2}\right\rangle \sim t^{\gamma}$
\cite{Bouchaud1990}. When $\gamma\neq1$,
this time dependence deviates from the linear function of time characteristic
for the Brownian motion. If $\gamma<1$, this phenomenon is called
subdiffusion. The occurrence of subdiffusion has been experimentally observed,
for example, in the behavior of individual colloidal particles in
random potential energy landscapes \cite{Evers2013}. Superdiffusion
($1<\gamma\leq2$) has been observed in vibrated granular media \cite{Scalliet2015}. 

Recently, in Refs.~\cite{Cherstvy2013,Cherstvy2013a,Cherstvy2014a,Cherstvy2014}
it has been suggested that the anomalous diffusion can be a result
of heterogeneous diffusion process (HDP), where the diffusion coefficient
depends on the position. Such spatially dependent diffusion can be
created in thermophoresis experiments using a local variation of the
temperature \cite{Maeda2012,Mast2013}. Theoretical aspects of a Brownian
particle diffusion in an environment with a position dependent temperature
has been investigated in \cite{Kazakevicius2015}.

The voter model started as a simple model of spatial competition between
two species \cite{Clifford1973,Liggett1999}, but over the decades
it has become one of the most studied models in opinion dynamics \cite{Castellano2009RevModPhys,Jedrzejewski2019CRP,Redner2019CRP}.
As the original model is extremely simple, involving only the recruitment
mechanism, it has seen a lot of modifications. Introduction of zealotry
\cite{Mobilia2007JStatMech,Khalil2018PRE}, random transitions \cite{Kirman1993QJE,Granovsky1995},
network topologies \cite{Alfarano2009Dyncon,Kononovicius2014EPJB,Carro2016,Peralta2018Chaos,Mori2019PRE,Gastner2019JPA}
and nonlinear interactions \cite{Artime2019CRPhys,Castellano2009PRE}
were shown to have effects on the phase behavior of the voter model.
Voter model has also seen applications in the modeling of electoral
and census data \cite{FernandezGarcia2014PRL,Sano2016,Kononovicius2017Complexity,Braha2017PlosOne,Kononovicius2019CompJStat}
and financial markets \cite{Alfarano2005CompEco,Alfarano2008Dyncon,Kononovicius2012,Gontis2014PlosOne,Franke2018,Kononovicius2019OB,Vilela2019PhysA}.
Anomalous diffusion, to the best of our knowledge, has not been considered
in the voter model or its modifications up until recently \cite{Kononovicius2020JStatMech}.
In \cite{Kononovicius2020JStatMech} it was shown that anomalous diffusion
can be observed by considering individual agent trajectories in a
modified voter model, thus providing an explanation for the observations
made in the parliamentary attendance data \cite{Vieira2019PRE}. Here
we show that after a nonlinear transformation of the observed variable
or the time scale the noisy voter model is able to exhibit subdiffusion,
superdiffusion, and localization phenomenon.

The paper is organized as follows. Section~\ref{sec:nvm} is dedicated
to the description of the original noisy voter model. Next, in Section~\ref{sec:observable}
we consider the nonlinear transformation of the observed variable.
In Section~\ref{sec:time-scale} we show that nonlinear transformation
of the time scale yields a similar process as the nonlinear transformation
of the observed variable.

\section{Noisy voter model}

\label{sec:nvm}

In \cite{Clifford1973} a simple model of competition between two
species was proposed in which a randomly selected member of species
replaced another member of the same or other species. In the context of
social system modeling, the core mechanism of this model is quite
similar to the recruitment mechanism. Actually, this model has first become
well known in opinion dynamics community, in which it is known as
the voter model \cite{Liggett1999,Castellano2009RevModPhys}.

Here we will consider modification of the original voter model, which
in the literature is known as the herding model \cite{Kirman1993QJE}
or the noisy voter model \cite{Granovsky1995}. While the original
model accounts only for the two particle interactions (one particle
adopts the state of the other particle), the noisy voter model also
accounts for one particle interactions (independent switching of the
state). Assuming that there are two states available and the number
of particles, $N$, is fixed, one can write the following transition
rates to describe the model:
\begin{align}
\pi\left(X\rightarrow X+1\right)=\pi^{+} & =\left(N-X\right)\left(r_{1}+hX\right),\nonumber \\
\pi\left(X\rightarrow X-1\right)=\pi^{-} & =X\left(r_{2}+h\left[N-X\right]\right).\label{eq:nvm-rates}
\end{align}
In the above $X$ is the number of particles in the first state, $r_{i}$
are the independent (one particle interaction) transition rates, while
$h$ is the recruitment (two particle interaction) transition rate.
Numerically this model can be simulated using the Gillespie method \cite{Gillespie1992AP}.
In our numerical simulations we use the Gillespie method with a modification,
which allows us to quickly obtain the results even with large $N$
(see Appendix~\ref{sec:simulation-method} for more details).

As the rates describe one step transitions, the birth--death process
framework can be used to derive a stochastic differential equation (abbr.
SDE) approximating the discrete process \cite{VanKampen2007NorthHolland}.
In the thermodynamic limit, $N\rightarrow\infty$, for $x=\frac{X}{N}$
the following SDE is obtained:
\begin{equation}
dx=\frac{\pi^{+}-\pi^{-}}{N}dt+\sqrt{\frac{\pi^{+}+\pi^{-}}{N^{2}}}dW\approx h\left[\varepsilon_{1}\left(1-x\right)-\varepsilon_{2}x\right]dt+\sqrt{2hx\left(1-x\right)}dW.\label{eq:sdex}
\end{equation}
In the above $\varepsilon_{i}=\frac{r_{i}}{h}$ are the relative independent
transition rates and $W$ is a standard Wiener process. Without loss of generality, we can set $h=1$ as
this parameter controls the base event rate of the process as a whole.

Steady state distribution of $x$ is Beta distribution, probability
density function (abbr. PDF) of which is given by:
\begin{equation}
P_{st}\left(x\right)=\frac{\Gamma\left(\varepsilon_{1}+\varepsilon_{2}\right)}{\Gamma\left(\varepsilon_{1}\right)\Gamma\left(\varepsilon_{2}\right)}x^{\varepsilon_{1}-1}\left(1-x\right)^{\varepsilon_{2}-1}.
\end{equation}
Beta distribution is observed in a variety of socio--economic scenarios.
Electoral \cite{Rigdon2009APR,Braha2017PlosOne,Kononovicius2017Complexity,Mori2019PRE},
religious adherence \cite{Ausloos2007EPL} and census data \cite{Kononovicius2019CompJStat}
seem to follow Beta distribution. While Beta distribution is not commonly
observed in finance, it was shown \cite{Ruseckas2011,Kononovicius2012}
that $x$ can be transformed into another variable, $y=\frac{x}{1-x}$,
which has a meaning of modulating (long--term) return. The new variable
$y$ has a power law steady state distribution and its time series
have long--range memory property, both of which are financial market
stylized facts \cite{Cont2001RQUF}. These empirical observations
indicate that the noisy voter model might be a reasonable model of opinion
dynamics in socio--economic scenarios.

\section{Transformation of the observed variable}

\label{sec:observable}

Here we will consider a more general form of nonlinear transformation
of $x$ described by SDE~\eqref{eq:sdex}:
\begin{equation}
y=\left(\frac{x}{1-x}\right)^{1/\alpha},
\end{equation}
with $\alpha\neq0$.
This transformation, with $\alpha=1$, was previously derived
in financial market context and was assumed to correspond to the long--term
varying component of return \cite{Ruseckas2011,Kononovicius2012}. The original
derivation of $ y $, see Section~3 in \cite{Kononovicius2012}, assumed
that excess demands of two different types of traders grow linearly with
the number of traders, but in general this dependence might be non--linear.
In a financial market scenario non--linear dependence might arise due to the
latent liquidity phenomenon \cite{Mastromatteo2014PRE,DallAmico2019JStat,
Lemhadri2020MML}, which arises because some traders hide their intentions
from other traders. With $\alpha > 1$, traders are assumed to hide their
intentions as the market goes out of equilibrium (excess demand grows). With
$\alpha < 1$, traders are assumed to hide their intentions as the market
approaches equilibrium (excess demand goes to zero).

Using Ito lemma for variable transformation, the following SDE for
$y$ is obtained:
\begin{equation}
dy=\frac{1}{\alpha^{2}}\left[\left(1+\alpha-\alpha\varepsilon_{2}\right)+\left(1-\alpha+\alpha\varepsilon_{1}\right)y^{-\alpha}\right]y\left(1+y^{\alpha}\right)dt+\sqrt{\frac{2}{\alpha^{2}}}y^{1-\frac{\alpha}{2}}\left(1+y^{\alpha}\right)dW.\label{eq:sdey}
\end{equation}
Note that SDE~\eqref{eq:sdex} is qualitatively invariant to $x\rightarrow1-x$
transformation. Namely, the transformed SDE would have exactly the
same form as SDE~\eqref{eq:sdex} with an exception that the parameters
$\varepsilon_{1}$ and $\varepsilon_{2}$ exchange their places. This
is a rather natural result as such transformation means simply relabeling
the states. Yet this has an important consequence on the SDE~\eqref{eq:sdey}:
this SDE is invariant to $y\rightarrow\frac{1}{y}$ transformation
with the same caveat. This property is particularly useful because
SDEs with the same $\left|\alpha\right|$ can be rearranged to have
the form given by Eq.~\eqref{eq:sdey}. Thus, without loss of generality
further in this section, we consider only $\alpha>0$ case.

Steady state PDF of $y$ is given by:
\begin{equation}
P_{st}\left(y\right)=\alpha\frac{\Gamma\left(\varepsilon_{1}+\varepsilon_{2}\right)}{\Gamma\left(\varepsilon_{1}\right)\Gamma\left(\varepsilon_{2}\right)}\cdot\frac{y^{\alpha\varepsilon_{1}-1}}{\left(1+y^{\alpha}\right)^{\varepsilon_{1}+\varepsilon_{2}}}.
\end{equation}
It is easy to see that PDF of $y$ has power law asymptotic behavior
for $y\gg1$: $P_{st}\left(y\right)\sim y^{-\alpha\varepsilon_{2}-1}$.
Raw steady state moments of $y$ are given by:
\begin{equation}
\left\langle y^{k}\right\rangle _{st}=\frac{\Gamma\left(\varepsilon_{1}+\frac{k}{\alpha}\right)\Gamma\left(\varepsilon_{2}-\frac{k}{\alpha}\right)}{\Gamma\left(\varepsilon_{1}\right)\Gamma\left(\varepsilon_{2}\right)}.\label{eq:sdey-stationary-moment}
\end{equation}
Note that for the $k$-th moment to exist $\alpha\varepsilon_{2}>k$
must hold. These describe the moments we expect to get as $t\rightarrow\infty$.
In what will soon follow we will derive an approximation for the time
evolution of the first two moments for finite $t$.

For convenience sake let us introduce the following notation:
\begin{align}
\eta_{\pm} & =1\pm\frac{\alpha}{2},\quad\lambda_{+}=1+\alpha\varepsilon_{2},\quad\lambda_{-}=1-\alpha\varepsilon_{1},\quad\mu=1+\frac{\alpha}{2}\left(\varepsilon_{1}-\varepsilon_{2}\right),\quad\sigma^{2}=\frac{2}{\alpha^{2}}.\label{eq:eta-params}
\end{align}
Then Eq.~\eqref{eq:sdey} can be rearranged into the following form:

\begin{equation}
dy=\sigma^{2}\left[\mu y+\left(\eta_{+}-\frac{\lambda_{+}}{2}\right)y^{2\eta_{+}-1}+\left(\eta_{-}-\frac{\lambda_{-}}{2}\right)y^{2\eta_{-}-1}\right]dt+\sigma\left(y^{\eta_{+}}+y^{\eta_{-}}\right)dW.\label{eq:sdey-eta}
\end{equation}
From this form it is evident that SDE for $y$ belongs to a general
class of SDEs exhibiting $1/f$ noise \cite{Ruseckas2011PhysRevE},
which is also known to exhibit anomalous diffusion \cite{Kazakevicius2016pre}.
For the results of \cite{Kazakevicius2016pre} to be applicable we
need to assume that $y\ll1$ or $y\gg1$ and approximate SDE~\eqref{eq:sdey-eta}
by:
\begin{align}
dy & =\sigma^{2}\Bigg[\mu y+\left(\eta_{-}-\frac{\lambda_{-}}{2}\right)y^{2\eta_{-}-1}\Bigg]dt+\sigma y^{\eta_{-}}dW,\quad\text{for}\:y\ll1,\\
dy & =\sigma^{2}\Bigg[\mu y+\left(\eta_{+}-\frac{\lambda_{+}}{2}\right)y^{2\eta_{+}-1}\Bigg]dt+\sigma y^{\eta_{+}}dW_{t},\quad\text{for}\:y\gg1.
\end{align}
In \cite{Kazakevicius2016pre} it was shown that the $\mu$ term influences
the moments of $y$ only for times close to the transition to a steady
state value. The $\mu$ term should not have a noticeable effect on
times when the anomalous diffusion is observed, thus let us ignore
it and use the following simplified SDE to obtain analytical approximations
for the temporal evolution of the first two moments:
\begin{equation}
dy=\sigma^{2}\left(\eta-\frac{\lambda}{2}\right)y^{2\eta-1}dt+\sigma y^{\eta}dW.\label{eq:sdey-simplest}
\end{equation}
In the above $\lambda$ is a parameter describing the additional drift
term. SDE~\eqref{eq:sdey-simplest} has been used to study the influence
of external potential on heterogeneous diffusion process \cite{Kazakevicius2015}.
The cases with $\lambda=0$, have been studied extensively in \cite{Cherstvy2013,Cherstvy2013a,Cherstvy2014a,Cherstvy2014}.

The time--dependent PDF of the process given by SDE~\eqref{eq:sdey-simplest}
was obtained in \cite{Kazakevicius2016pre}: 
\begin{equation}
P\left(y,t\right|\left.y_{0},0\right)=\frac{y^{\frac{1-2\eta-\lambda}{2}}y_{0}^{\frac{1-2\eta+\lambda}{2}}}{\left|\eta-1\right|\sigma^{2}t}\exp\left(-\frac{y^{2(1-\eta)}+y_{0}^{2(1-\eta)}}{2(\eta-1)^{2}\sigma^{2}t}\right)I_{\frac{\lambda+1-2\eta}{2(\eta-1)}}\left(\frac{y^{(1-\eta)}y_{0}^{(1-\eta)}}{(\eta-1)^{2}\sigma^{2}t}\right).\label{eq:time-pdf-y}
\end{equation}
Here $I_{n}\left(\dots\right)$ is the modified Bessel function of
the first kind of order $n$. The time--dependent PDF satisfies the
initial condition $P\left(y,t\right|\left.y_{0},0\right)=\delta\left(y-y_{0}\right)$
(with $y_{0}$ being the initial value). To ensure absorption at the
boundaries $\frac{\lambda+1-2\eta}{2(\eta-1)}>-1$ must hold. This
condition holds whenever $\lambda>1$ and $\eta>1$ or $\lambda<1$
and $\eta<1$. By looking at Eq.~\eqref{eq:eta-params} it is evident
that one of these conditions is always satisfied as the relative individual
transition rates, $\varepsilon_{i}$, are always positive (due to
this one pair of $\lambda$ and $\eta$ is always larger than $1$,
while the other is always smaller). In this regard steady state moment
existence condition is a bit more restrictive.

\begin{figure}
\begin{centering}
\includegraphics[width=0.9\textwidth]{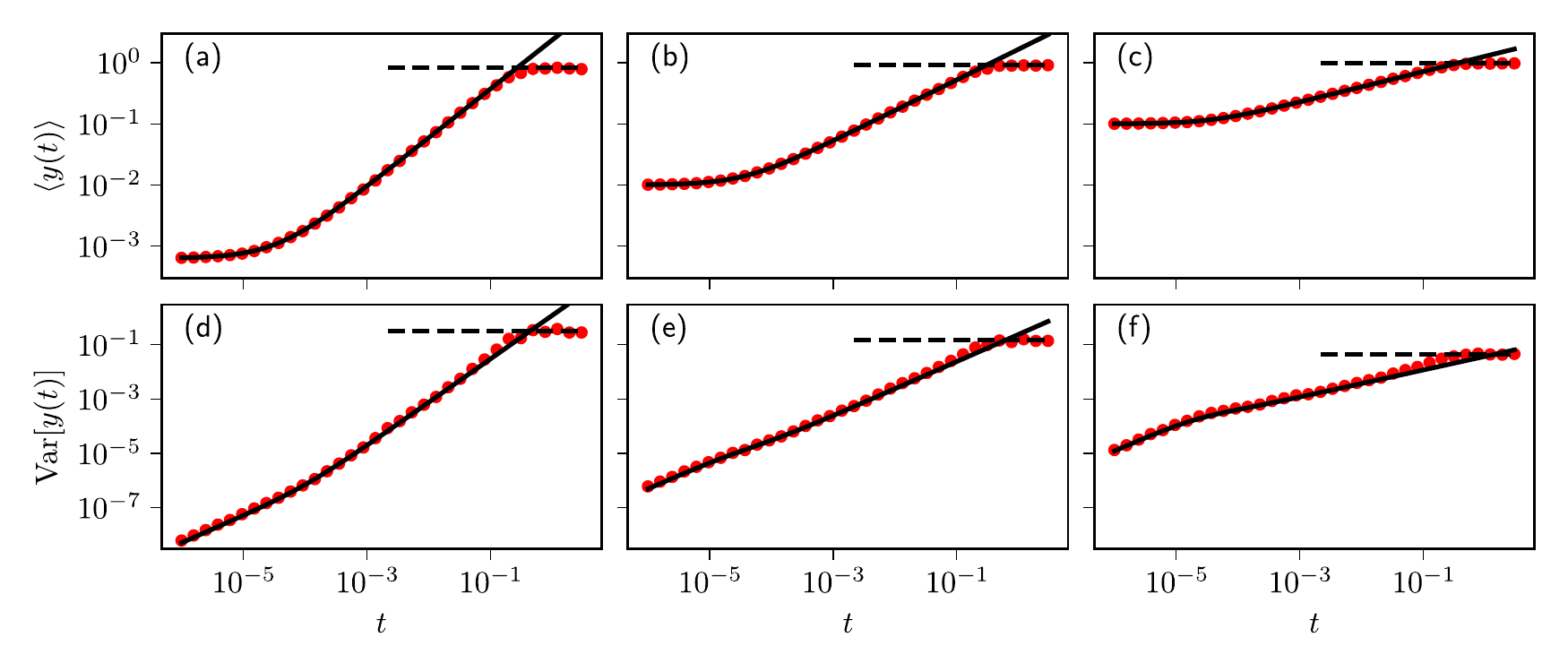}
\par\end{centering}
\caption{First two time--dependent moments of the noisy voter model with transformed
observable for $\mu=0$ and $y_{0}\ll1$ case: numerical results from
the discrete model (red circles), analytical approximation derived
from Eq.~\eqref{eq:sdey-time-moments} (solid black curve) and stationary
value derived from Eq.~\eqref{eq:sdey-stationary-moment} (dashed
black curve). Discrete model parameters: $N=10^{5}$, $X_{0}=10$,
$\varepsilon_{1}=3$, $\varepsilon_{2}=\varepsilon_{1}+\frac{2}{\alpha}$
(all cases), $\alpha=1.25$ ((a), (d)), $2$ ((b), (e)) and $4$ ((c),
(f)). Respective SDE~\eqref{eq:sdey-simplest} parameters: $y_{0}=9999^{-\frac{1}{\alpha}}$,
$\eta=1-\frac{\alpha}{2}$, $\lambda=1-3\alpha$, $\sigma^{2}=\frac{2}{\alpha^{2}}$,
$\mu=0$.\label{fig:mu-zero-low}}
\end{figure}

\begin{figure}
\begin{centering}
\includegraphics[width=0.9\textwidth]{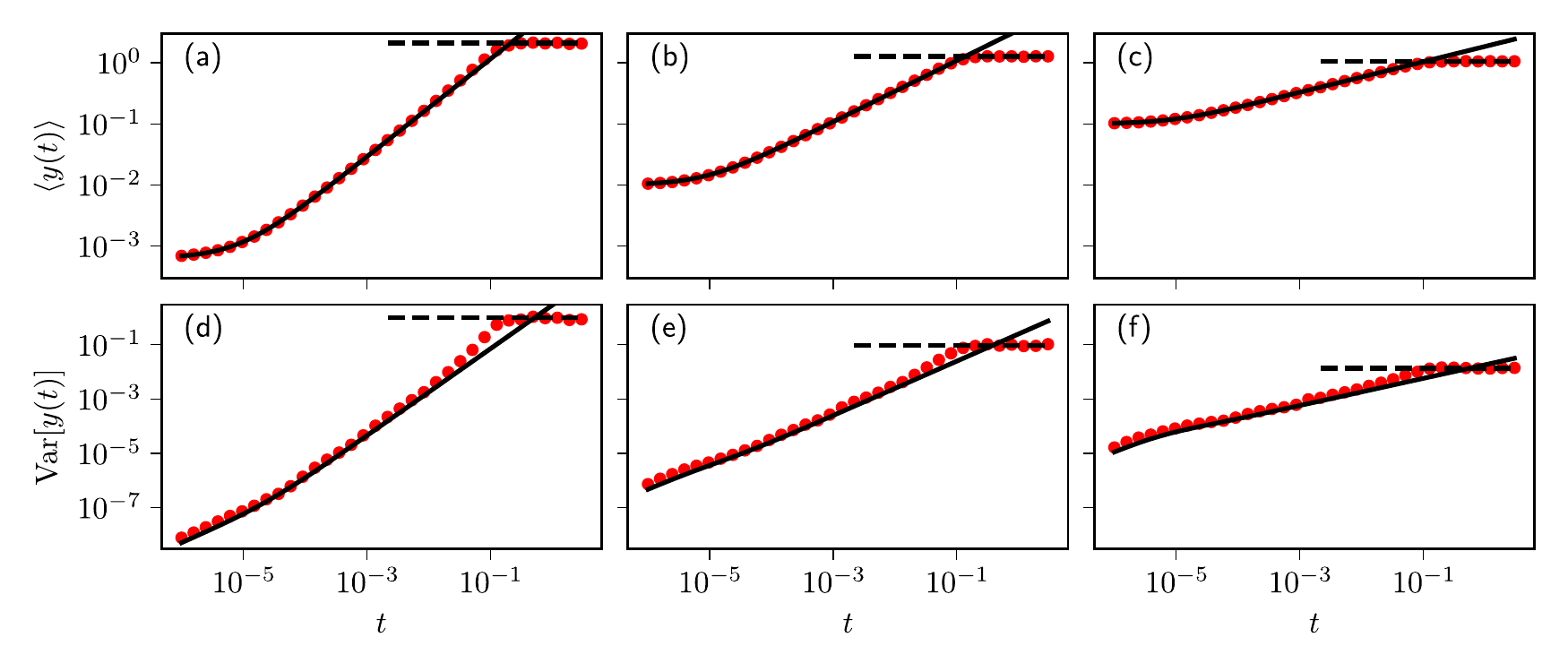}
\par\end{centering}
\caption{First two time--dependent moments of the noisy voter model with transformed
observable for $\mu>0$ and $y_{0}\ll1$ case: numerical results from
the discrete model (red circles), analytical approximation derived
from Eq.~\eqref{eq:sdey-time-moments} (solid black curve) and stationary
value derived from Eq.~\eqref{eq:sdey-stationary-moment} (dashed
black curve). Discrete model parameters: $N=10^{5}$, $X_{0}=10$,
$\varepsilon_{1}=12$, $\varepsilon_{2}=\varepsilon_{1}-\frac{8}{\alpha}$
(all cases), $\alpha=1.25$ ((a), (d)), $2$ ((b), (e)) and $4$ ((c),
(f)). Respective SDE~\eqref{eq:sdey-simplest} parameters: $y_{0}=9999^{-\frac{1}{\alpha}}$,
$\eta=1-\frac{\alpha}{2}$, $\lambda=1-12\alpha$, $\sigma^{2}=\frac{2}{\alpha^{2}}$,
$\mu=5$.\label{fig:mu-pos-low}}
\end{figure}

\begin{figure}
\begin{centering}
\includegraphics[width=0.9\textwidth]{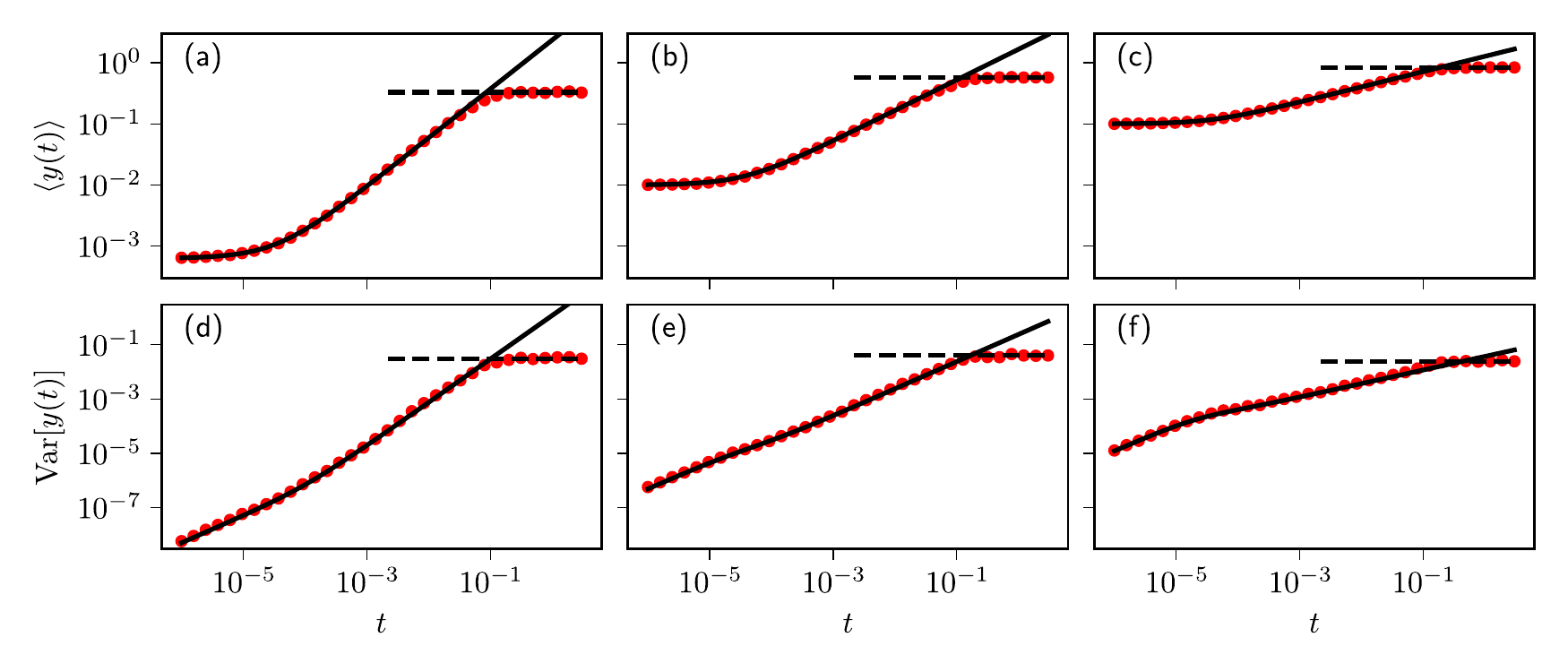}
\par\end{centering}
\caption{First two time--dependent moments of the noisy voter model with transformed
observable for $\mu<0$ and $y_{0}\ll1$ case: numerical results from
the discrete model (red circles), analytical approximation derived
from Eq.~\eqref{eq:sdey-time-moments} (solid black curve) and stationary
value derived from Eq.~\eqref{eq:sdey-stationary-moment} (dashed
black curve). Discrete model parameters: $N=10^{5}$, $X_{0}=10$,
$\varepsilon_{1}=3$, $\varepsilon_{2}=\varepsilon_{1}-\frac{8}{\alpha}$
(all cases), $\alpha=1.25$ ((a), (d)), $2$ ((b), (e)) and $4$ ((c),
(f)). Respective SDE parameters: $y_{0}=9999^{-\frac{1}{\alpha}}$,
$\eta=1-\frac{\alpha}{2}$, $\lambda=1-3\alpha$, $\sigma^{2}=\frac{2}{\alpha^{2}}$,
$\mu=-5$.\label{fig:mu-neg-low}}
\end{figure}

\begin{figure}
\begin{centering}
\includegraphics[width=0.9\textwidth]{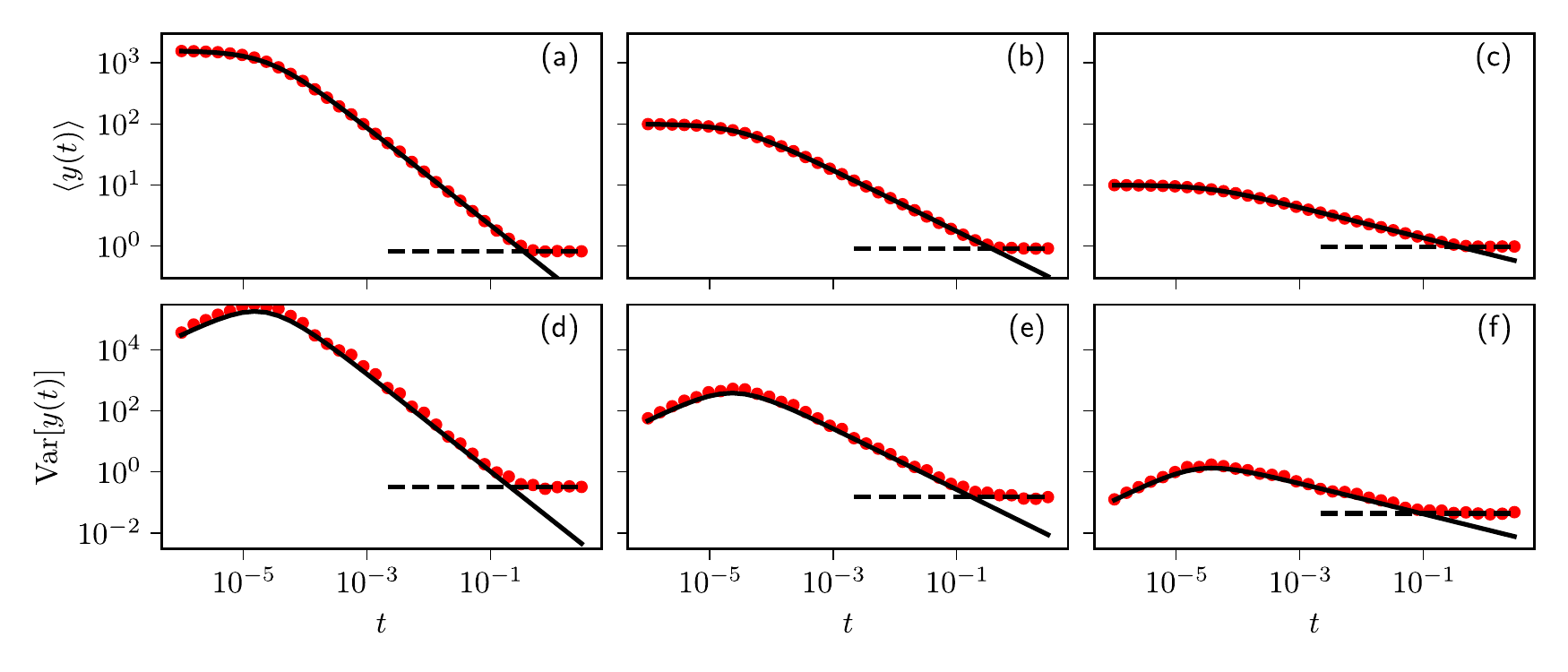}
\par\end{centering}
\caption{First two time--dependent moments of the noisy voter model with transformed
observable for $\mu=0$ and $y_{0}\gg1$ case: numerical results from
the discrete model (red circles), analytical approximation derived
from Eq.~\eqref{eq:sdey-time-moments} (solid black curve) and stationary
value derived from Eq.~\eqref{eq:sdey-stationary-moment} (dashed
black curve). Discrete model parameters: $N=10^{5}$, $X_{0}=N-10$,
$\varepsilon_{1}=3$, $\varepsilon_{2}=\varepsilon_{1}+\frac{2}{\alpha}$
(all cases), $\alpha=1.25$ ((a), (d)), $2$ ((b), (e)) and $4$ ((c),
(f)). Respective SDE~\eqref{eq:sdey-simplest} parameters: $y_{0}=9999^{\frac{1}{\alpha}}$,
$\eta=1+\frac{\alpha}{2}$, $\lambda=3+3\alpha$, $\sigma^{2}=\frac{2}{\alpha^{2}}$,
$\mu=0$.\label{fig:mu-zero-high}}
\end{figure}

\begin{figure}
\begin{centering}
\includegraphics[width=0.9\textwidth]{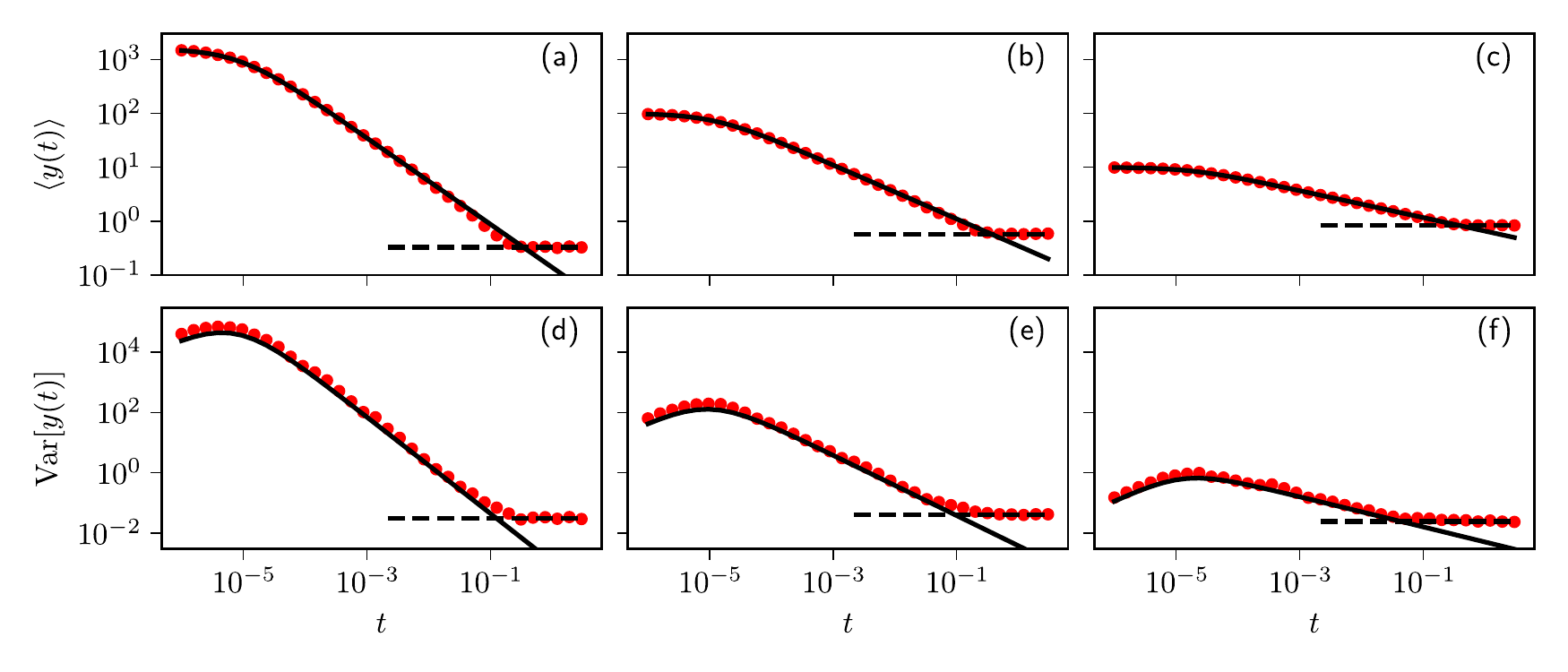}
\par\end{centering}
\caption{First two time--dependent moments of the noisy voter model with transformed
observable for $\mu<0$ and $y_{0}\gg1$ case: numerical results from
the discrete model (red circles), analytical approximation derived
from Eq.~\eqref{eq:sdey-time-moments} (solid black curve) and stationary
value derived from Eq.~\eqref{eq:sdey-stationary-moment} (dashed
black curve). Discrete model parameters: $N=10^{5}$, $X_{0}=N-10$,
$\varepsilon_{1}=3$, $\varepsilon_{2}=\varepsilon_{1}-\frac{8}{\alpha}$
(all cases), $\alpha=1.25$ ((a), (d)), $2$ ((b), (e)) and $4$ ((c),
(f)). Respective SDE~\eqref{eq:sdey-simplest} parameters: $y_{0}=9999^{\frac{1}{\alpha}}$,
$\eta=1+\frac{\alpha}{2}$, $\lambda=3\alpha-7$, $\sigma^{2}=\frac{2}{\alpha^{2}}$,
$\mu=-5$.\label{fig:mu-neg-high}}
\end{figure}

From Eq.~\eqref{eq:time-pdf-y} we can calculate the $k$-th time--dependent
raw moment of $y$:
\begin{eqnarray}
\left\langle y^{k}\left(t,y_{0}\right)\right\rangle  & = & \int_{0}^{\infty}y^{k}P\left(y,t\right|\left.y_{0},0\right)dy=\nonumber \\
 & = & \frac{\Gamma\left(\frac{\lambda-1-k}{2(\eta-1)}\right)}{\Gamma\left(\frac{\lambda-1}{2(\eta-1)}\right)}\left(2(\eta-1)^{2}\sigma^{2}t\right)^{\frac{k}{2(1-\eta)}}{}_{1}F_{1}\left(\frac{k}{2(\eta-1)};\frac{\lambda-1}{2(\eta-1)};-\frac{y_{0}^{2(1-\eta)}}{2(\eta-1)^{2}\sigma^{2}t}\right).\label{eq:sdey-time-moments}
\end{eqnarray}
In the above $_{1}F_{1}\left(\dots\right)$ is the Kummer confluent
hypergeometric function. The time--dependent moments will be finite
as long as respective steady state moments exist. For the intermediate
times \cite{Kazakevicius2016pre}, 
\begin{equation}
\frac{y_{0}^{2(1-\eta)}}{2(\eta-1)^{2}\sigma^{2}}\ll t,\label{eq:sdey-times}
\end{equation}
the hypergeometric function of Eq.~\eqref{eq:sdey-time-moments}
is approximately equal to $1$. Thus, for the intermediate times the
time--dependent raw moments will grow as a power law function of
time: 
\begin{equation}
\left\langle y^{k}\left(t,y_{0}\right)\right\rangle \approx\left\langle y^{k}\left(t\right)\right\rangle \sim t^{\gamma}.
\end{equation}
Here $\gamma=\frac{k}{2(1-\eta)}$. Therefore, by manipulating $\eta$,
or the power of transformation $\alpha$, we can select model parameters
so that we would observe subdiffusion, normal diffusion, superdiffusion
or localization. From Eq.~\eqref{eq:sdey-time-moments} follows that
the sign of power law exponent $\gamma$ depends on whether initial
condition satisfies inequality $y_{0}\ll1$ ($\gamma>0$ ) or $y_{0}\gg1$
($\gamma<0$). In Figs.~\ref{fig:mu-zero-low}--\ref{fig:mu-neg-low}
we have set $y_{0}\ll1$ and selected values of $\alpha$ specifically
to show the three diffusive behaviors: $\alpha=1.25$ for superdiffusion
($\gamma=1.6$), $\alpha=2$ for normal diffusion ($\gamma=1$) and
$\alpha=4$ for subdiffusion ($\gamma=0.5$). In Figs.~\ref{fig:mu-zero-high}--\ref{fig:mu-neg-high}
we have set $y_{0}\gg1$ with the same $\alpha$ values to show different
inverse power law diffusive behaviors with exponents $\gamma=-1.6;-1;-0.5$
respectively to $\alpha=1.25;2;4$. In previous figures observed variance
slow return to steady state is called localization ant it is also
a property of heterogeneous diffusion process \cite{Cherstvy2014}.
So, we can conclude that nonlinear transformations of noisy voter model
exhibit the same diffusive properties as heterogeneous diffusion
process for intermediate times. 

Note that analytical approximations derived from Eq.~\eqref{eq:sdey-time-moments}
are rather good up to the larger times where the steady state behavior
takes over. The analytical approximations hold reasonably well even
for $\mu\neq0$, though the disagreement between numerical and analytical
results around larger times is a bit more apparent in anomalous variance
growth (see Figs.~\ref{fig:mu-zero-low} and \ref{fig:mu-pos-low}
d,e,f ).

\begin{figure}
\begin{centering}
\includegraphics[width=0.9\textwidth]{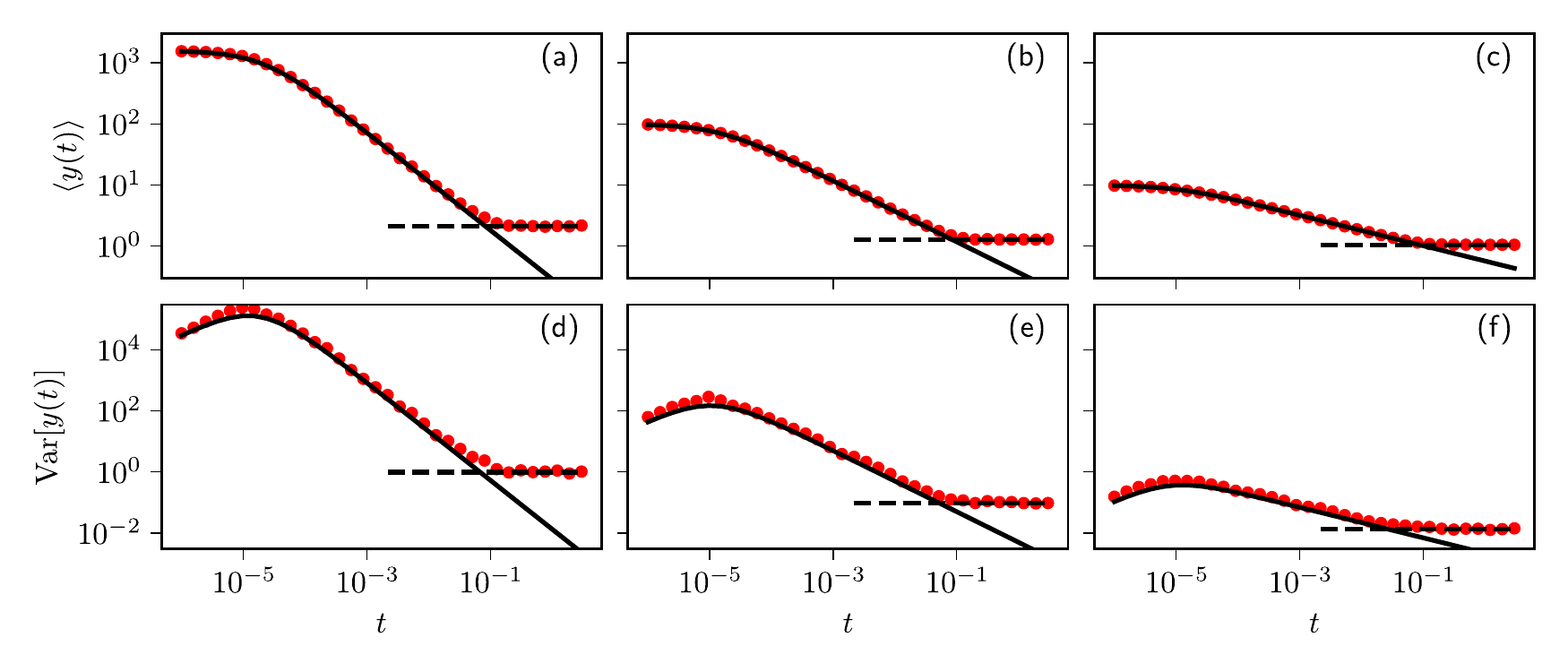}
\par\end{centering}
\caption{First two time--dependent moments of the noisy voter model with transformed
observable for $\mu>0$ and $y_{0}\gg1$ case: numerical results from
the discrete model (red circles), analytical approximation derived
from Eq.~\eqref{eq:sdey-time-moments} (solid black curve) and stationary
value derived from Eq.~\eqref{eq:sdey-stationary-moment} (dashed
black curve). Discrete model parameters: $N=10^{5}$, $X_{0}=N-10$,
$\varepsilon_{1}=12$, $\varepsilon_{2}=\varepsilon_{1}-\frac{8}{\alpha}$
(all cases), $\alpha=1.25$ ((a), (d)), $2$ ((b), (e)) and $4$ ((c),
(f)). Respective SDE~\eqref{eq:sdey-simplest} parameters: $y_{0}=9999^{\frac{1}{\alpha}}$,
$\eta=1+\frac{\alpha}{2}$, $\lambda=3\alpha-7$, $\sigma^{2}=\frac{2}{\alpha^{2}}$,
$\mu=5$.}
\end{figure}

\section{Transformation of the time scale}

\label{sec:time-scale}

In Refs.~\cite{Ruseckas2011,Kononovicius2012} variable event time
scale was introduced into the SDE of the noisy voter model by the
means of $\tau\left(x\right)$ function.Such inclusion of the
variable event time scale stems from empirical observation in the
financial data \cite{Rak2013APP,Gontis2014PlosOne}: absolute returns
correlate most with square of the trading volume. Coincidentally such
an assumption allows easy reproduction of long-range memory phenomenon
across a broad range of frequencies
\cite{Kononovicius2012,Ruseckas2011,Gontis2014PlosOne}.
We include $\tau\left(x\right)$ into SDE~\eqref{eq:sdex}
by replacing base event rate of the process $h$ by $1/\tau(x)$:
\begin{equation}
dx=\left[\varepsilon_{1}\left(1-x\right)-\varepsilon_{2}x\right]\frac{dt}{\tau\left(x\right)}+\sqrt{\frac{2x\left(1-x\right)}{\tau\left(x\right)}}dW.\label{eq:sdex-tau}
\end{equation}
Based on the aforementioned empirical observation and the
experience reproducing long-range memory phenomenon, let us consider
the following power-law form of $\tau \left( x \right) $:
\begin{equation}
\tau\left(x\right)=x^{1-2\eta}.
\end{equation}
Then the SDE~\eqref{eq:sdex-tau} would take the following form 
\begin{equation}
dx=\left[\varepsilon_{1}\left(1-x\right)-\varepsilon_{2}x\right]x^{2\eta-1}dt+\sqrt{2x^{2\eta}\left(1-x\right)}dW.\label{eq:sdex-tau-eta}
\end{equation}
Note that SDE~\eqref{eq:sdex-tau-eta} can be interpreted as noisy
voter model interactions occurring in the internal time $\tau$, which
is related to the physical time $t$ by the means of transformation
$dt=x\left(\tau\right)^{1-2\eta}d\tau$ (for more details see Appendix~\eqref{sec:app-time-nvm}).

Steady state distribution of $x$ is still Beta distribution, though
its parameters are slightly different:
\begin{equation}
P_{st}(x)=\frac{\Gamma(\varepsilon_{\eta}+\varepsilon_{2})}{\Gamma(\varepsilon_{\eta})\Gamma(\varepsilon_{2})}x^{\varepsilon_{\eta}-1}x^{\varepsilon_{2}-1},
\end{equation}
where $\varepsilon_{\eta}=\varepsilon_{1}+1-2\eta$. Raw steady state
moments of $x$ are given by:
\begin{equation}
\left\langle x^{k}\right\rangle _{st}=\frac{\Gamma\left(\varepsilon_{\eta}+\varepsilon_{2}\right)\Gamma\left(\varepsilon_{\eta}+k\right)}{\Gamma\left(\varepsilon_{\eta}\right)\Gamma\left(\varepsilon_{\eta}+\varepsilon_{2}+k\right)}.\label{eq:sdex-stationary-moments}
\end{equation}

\begin{figure}
\begin{centering}
\includegraphics[width=0.9\textwidth]{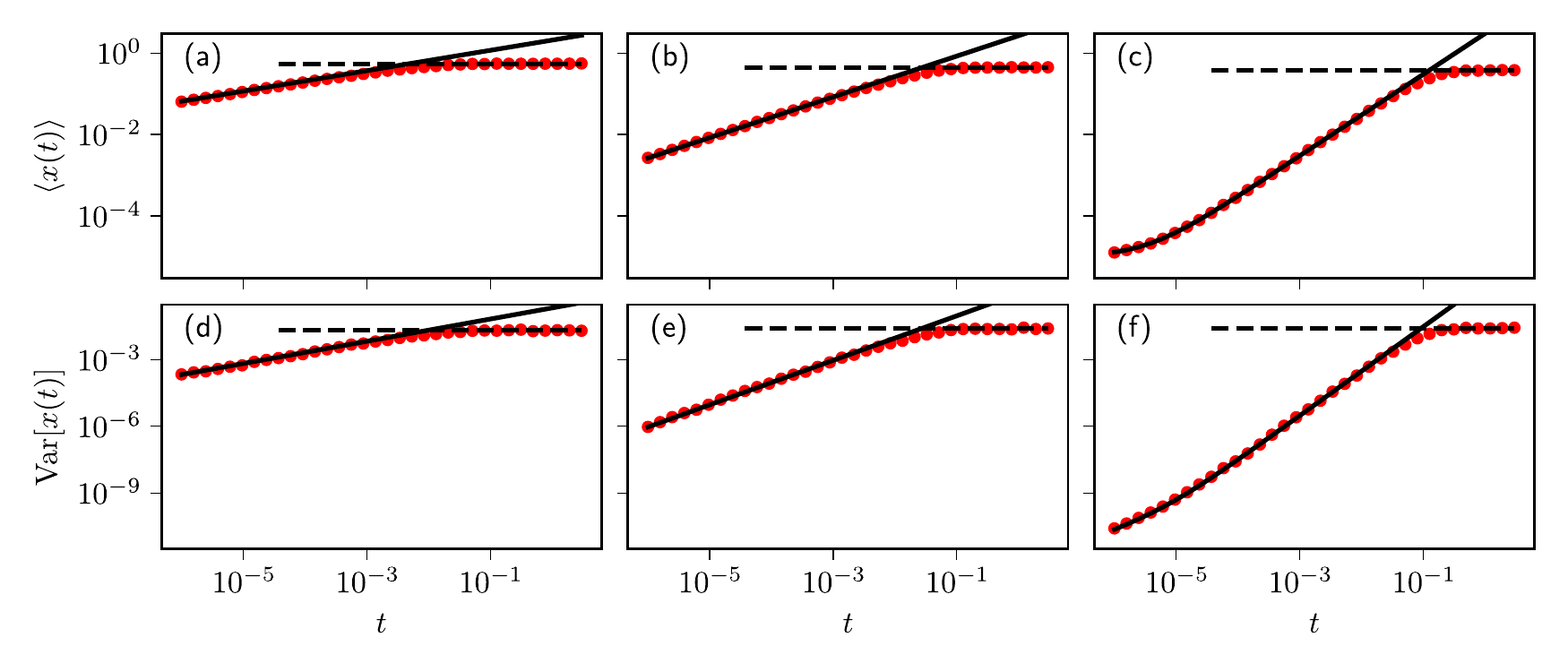}
\par\end{centering}
\caption{First two time--dependent moments of the noisy voter model with transformed
time scale: numerical results from the discrete model (red circles),
analytical approximation derived from Eq.~\eqref{eq:sdex-time-moments}
(solid black curve) and stationary value derived from Eq.~\eqref{eq:sdex-stationary-moments}
(dashed black curve). Model parameters: $N=10^{6}$, $X_{0}=10$,
$\varepsilon_{1}=3$, $\varepsilon_{2}=5$ (all cases), $\eta=-1$
((a), (d)), $0$ ((b), (e)) and $0.5$ ((c), (f)). Respective SDE~\eqref{eq:sdex-eta-approx}
parameters: $\lambda=2\eta-3$.\label{fig:tau}}
\end{figure}

Let us introduce the following notation:
\begin{equation}
\lambda=2\eta-\varepsilon_{1},\quad x_{c}=\frac{\varepsilon_{1}}{\varepsilon_{1}+\varepsilon_{2}}.
\end{equation}
Using this notation, we can rewrite SDE~\eqref{eq:sdex-tau-eta} as:
\begin{equation}
dx=2\left(\eta-\frac{\lambda}{2}\right)x^{2\eta-1}\Bigg[1-\frac{x}{x_{c}}\Bigg]dt+x^{\eta}\sqrt{2\left(1-x\right)}dW.
\end{equation}
If $x$ is sufficiently small, $x\ll x_{c}$, then we can approximate
SDE~\eqref{eq:sdex-tau-eta} by: 
\begin{equation}
dx=2\left(\eta-\frac{\lambda}{2}\right)x^{2\eta-1}dt+\sqrt{2}x^{\eta}dW.\label{eq:sdex-eta-approx}
\end{equation}
This SDE has identical form as SDE~\eqref{eq:sdey-simplest}, which
we have considered in a previous section.  Therefore, given that initial
value $x_{0}$ is much smaller than $x_{c}$, we can use the result
from previous section, Eq.~\eqref{eq:sdey-time-moments}, to approximate
the $k$-th time--dependent raw moment of $x$ as follows:
\begin{eqnarray}
\left\langle x^{k}\left(t,x_{0}\right)\right\rangle  & = & \frac{\Gamma\left(\frac{\lambda-1-k}{2(\eta-1)}\right)}{\Gamma\left(\frac{\lambda-1}{2(\eta-1)}\right)}\left(4(\eta-1)^{2}t\right)^{\frac{k}{2(1-\eta)}}{}_{1}F_{1}\left(\frac{k}{2(\eta-1)};\frac{\lambda-1}{2(\eta-1)};-\frac{x_{0}^{2(1-\eta)}}{4(\eta-1)^{2}t}\right).\label{eq:sdex-time-moments}
\end{eqnarray}
As previously, the asymptotic behavior of the moments for intermediate
times will be power law function, $t^{\gamma}$, with exponent $\gamma=\frac{k}{2(1-\eta)}$.
Consequently, for variance ($k=2$) we will observe superdiffusion
when $0<\eta\leq0.5$ (Fig.~\ref{fig:tau}d) and subdiffusion when
$\eta<0$ (Fig.~\ref{fig:tau}d). Note that the unmodified noisy voter
model corresponds to $\eta=0.5$ and the ballistic regime is expected
to be observed. In Fig.~\ref{fig:tau} we have shown that the proposed
approximation works well until steady state behavior takes over.

\section{Conclusions}

\label{sec:concl}

In this paper, we have considered nonlinear transformations of the
noisy voter model. We have shown that transformations of both observation
variable and time scale lead to a process, which is able to reproduce
various regimes of diffusion: subdiffusion, normal diffusion, and
superdiffusion. In the case of observation variable transformation, we have also
observed localization phenomenon for the intermediate times. To the best of
our knowledge anomalous diffusion was not previously examined in the
context of the voter models.

Nonlinear transformations we consider here are related to the
latent liquidity phenomenon observed in the financial markets
\cite{Mastromatteo2014PRE,DallAmico2019JStat,Lemhadri2020MML}. This
phenomenon arises because some traders hide their intentions from their
peers. Hence nonlinear transformation parameters could serve as a measure
of latent liquidity, but the proper characterization of anomalous diffusion
of single trajectories still eludes the scientific community
\cite{Munoz2020SPIE}.

We have derived analytical approximations for the temporal evolution
of the raw moments. Using these approximations one can easily approximate
the temporal evolution of mean, variance, and other higher moments.
Our derivations were based on the approximation of the time--dependent
PDF of the processes, which could be used to derive first passage
time distribution or to do Bayesian inference using the noisy voter
model. As far as it is known to us, there is no closed form expression
for the time--dependent PDF of the noisy voter model. There exists
only expression involving infinite sum \cite{Lanska1994}.

\section*{Author contributions}

Conceptualization: R.K.; Methodology: R.K. and A.K.; Software: A.K.;
Writing -- Original Draft: R.K. and A.K.; Writing -- Review \& Editing: R.K.
and A.K.; Visualization: A.K.

\appendix

\section{Simulation method}

\label{sec:simulation-method}

In this section, we briefly discuss the numerical simulation method used
in this paper.

Our approach to numerical simulation relies on the Gillespie
method \cite{Gillespie1992AP}. Originally Gillespie method was developed to
generate statistically correct discrete trajectories of chemical reactions
involving finite number of individual molecules.
This method relies on a fact that sojourn times between the chemical reactions
are exponentially distributed and independent. Therefore instead of
considering sojourn times of every individual reaction, we can consider
only the shortest sojourn time (of the reaction that will happen first).
Due to the nature of the order statistics of an exponential distribution,
instead of sampling sojourn times for each individual reaction, we can
sample just the shortest sojourn time.

In the noisy voter model there are just two possible types of
events, either $X\rightarrow X+1$ or $X\rightarrow X-1$. To model either
of the events happening we sample sojourn time from an exponential
distribution whose rate is given by a sum of the both event rates,
Eq.~\eqref{eq:nvm-rates}:
\begin{equation}
\Delta t_{i}\sim\mathrm{Exp}\left(\pi^{+}+\pi^{-}\right).
\end{equation}
After $\Delta t_{i}$ time, passes $X$ is incremented with probability
$p^{+}=\frac{\pi^{+}}{\pi^{+} + \pi^{-}}$, otherwise
(with complementary probability $ p^{-} = 1 - p^{+} $) $ X $ is decremented.
After some event happens the rates are updated and new sojourn time can be
generated to further sample the trajectory.

To get value of $X$ at time $t$ we simply run the model until the
internal clock of the model passes $t$. As the model runs, we keep both the
new and the old value of $X$. As soon as the internal clock passes $t$
we return the old value of $X$ as $X\left(t\right)$.

With small $N$ the numerical simulation is rather quick, but in general
its time complexity scales as $N^{2}$, therefore for larger $N$
the simulations become too long. Especially as we need $10^{3}$ runs
for each of the parameter sets. On our hardware simulations with $N=10^{5}$
ran for a few hours (using a single parameter set). Yet in certain cases,
we needed $N=10^{6}$ for the anomalous diffusion to be properly observable
and not being distorted by the discreteness effects. Simulations with
$N=10^{6}$ would take a couple of weeks to complete for a single parameter
set.

Yet we can improve the speed of simulations by noticing that time
complexity is actually $X\cdot\left(N-X\right)$. In other words,
simulation proceeds faster when $X$ is close $0$ or $N$ than when
$X$ is intermediate. Also, we are mostly interested in the behavior
when $X$ is close to these boundaries and we care less about the
intermediate values (as they correspond to the steady state behavior).
Therefore, we can decrease $N$ as $X$ approaches the intermediate
values. Whenever we decrease $N$ we keep the ratio $\frac{X}{N}$
constant. In our implementation, we divide $X$ and $N$ by $10$ when
threshold, $X=100$ or $X=N-100$, is reached, we do so until $N$
drops to $10^{3}$. The threshold and divisor were selected arbitrarily
as for these values we feel that such resolution is enough for the
discreteness effects not to be too obvious.

Note that we have implemented only the downscaling of $N$, so our
modification works when $\varepsilon_{i}>1$. This limitation is not
important in the scope of this paper, as for the first two steady
state moments to exist we need to have $\varepsilon_{i}>2$. To allow for
any positive $\varepsilon_{i}$ one should also implement the upscaling
of $N$.

The modified implementation runs orders of magnitude faster than the
pure Gillespie implementation. For $N=10^{6}$ the simulation runs
not for weeks, but for a couple of minutes. The improvement in speed
would be even more noticeable for even larger $N$.

To make sure that our implementation produces correct results we have
run a single batch of simulations with $N=10^{5}$ and compared the
results. The differences were too small to be actually visible.

We have made our code, written using Python, C, and Shell, available
from GitHub (URL: \href{https://github.com/akononovicius/anomalous-diffusion-in-nonlinear-transformations-of-the-noisy-voter-model}{https://github.com/akononovicius/anomalous-diffusion-in-nonlinear-transformations-of-the-noisy-voter-model}).
Some of our Shell scripts rely on GNU parallel utility \cite{Tange2018},
which allows us to run multiple simulations at once.

\section{Time transformation of the noisy voter model}

\label{sec:app-time-nvm}

In this section we derive SDE~\eqref{eq:sdex-tau-eta}, which describes
the noisy voter model with variable event time scale, from the time
subordinated original noisy voter model described by SDE~\eqref{eq:sdex}.

In general, stochastic process in operational time $\tau$ is described
by
\begin{equation}
dx_{\tau}=a\left(x_{\tau}\right)d\tau+b\left(x_{\tau}\right)dW_{\tau}.\label{eq:opertional-time}
\end{equation}
Here we assume that the small increments of the physical time $t$
are deterministic and are proportional to the increments of the operational
time $\tau$. Therefore, the physical time $t$ is related to the
internal time $\tau$ via the equation 
\begin{equation}
dt_{\tau}=g\left(x_{\tau}\right)d\tau.\label{eq:subord}
\end{equation}
The positive function $g\left(x\right)$ is the intensity of random
time that depends on the intensity of the stochastic process $x$.
In general, the relationship between the physical time $t$ and the
operational time $\tau$ can be nondeterministic. Namely, Eq.~\eqref{eq:subord}
can also include a stochastic term \cite{Kazakevicius2015physA}.
Such a procedure is commonly known as subordination. 

To understand Eq.~\eqref{eq:subord} influence on the stochastic
process $x$ it is useful to interpret Eq.~\eqref{eq:opertional-time}
as describing the diffusion of a particle in a nonhomogeneous medium.
If so, then the function $g\left(x\right)$ models the position of
the structures responsible for either slowing down or accelerating
the particle. Large values of $g\left(x\right)$ corresponds to slowing
the particle and small $g\left(x\right)$ leads to the acceleration
of the diffusion. 

In \cite{Ruseckas2015} it was shown that corresponding SDE~\eqref{eq:sdex}
the SDE describing the stochastic process $x$ in physical time is
\begin{equation}
dx_{t}=\frac{a(x_{t})}{g(x_{t})}d\tau+\frac{b(x_{t})}{\sqrt{g(x_{t})}}dW_{t}.\label{eq:time-trans-gen}
\end{equation}

If we assume that underlying stochastic process in operational time
$\tau$ is governed by the noisy voter model \eqref{eq:sdex} from
\eqref{eq:time-trans-gen} follows that
\begin{equation}
dx_{t}=\Bigg[\varepsilon_{1}\left(1-x_{t}\right)-\varepsilon_{2}x_{t}\Bigg]\frac{d\tau}{g\left(x_{t}\right)}+\sqrt{\frac{2x_{t}\left(1-x_{t}\right)}{g\left(x_{t}\right)}}dW_{t}.\label{eq:sdex-subord}
\end{equation}
If we assume that an intensity of random time is power law function
$g\left(x\right)=x^{1-2\eta}$ we obtain the same SDE~\eqref{eq:sdex-tau-eta}.

It is straightforward to include $g\left(x\right)$ into the noisy
voter model event rates, Eq.~\eqref{eq:nvm-rates}. As this function
changes the speed of diffusion, it scales the event rates:
\begin{align}
\pi^{+} & =\frac{1}{g\left(\frac{X}{N}\right)}\left(N-X\right)\left(\sigma_{1}+hX\right),\nonumber \\
\pi^{-} & =\frac{1}{g\left(\frac{X}{N}\right)}\left(\sigma_{2}+h\left[N-X\right]\right).
\end{align}
By treating the rates as in Eq.~\eqref{eq:sdex}, we get SDE identical
to SDE~\eqref{eq:sdex-subord}. By selecting the power law form of
$g\left(x\right)$ we obtain the same SDE~\eqref{eq:sdex-tau-eta}.

\end{document}